# Picometer-Scale Spatial Symmetry Breaking in Active Transmissive Metasurfaces

Martin Thomaschewski, Ruzan Sokhoyan, Elisabetta Schneider, and Harry Atwater*

*Thomas J. Watson Laboratory of Applied Physics, California Institute of Technology, Pasadena, CA 91125, USA*

E-mail: haa@caltech.edu

## Abstract

Active transmissive metasurfaces are central building blocks for future compact, cascadable optical systems, enabling the stacking of multiple functional layers for advanced dynamic beam shaping, photonic neural networks, depth sensing, and holography. We present a transmissive electro-optic metasurface based on silicon-on-lithium-niobate, where an array of silicon waveguides with periodic perturbations, individually controlled at the 100 pm scale, supports well-defined high-Q (>2000) guided-mode resonances (GMRs). We incorporate interdigitated push-pull electrodes between subwavelength-spaced GMR elements to locally tune the refractive index in the lithium niobate substrate, thereby shifting the GMR resonance and enabling opposite phase and amplitude modulation between neighboring radiative elements. In a geometrically symmetric metasurface, this effect introduces electro-optic beam splitting via diffraction, with diffraction efficiencies as high as 3%. By introducing controlled passive resonance detuning via 100 pm scale perturbation shifts, we realize a Vernier-type enhancement mechanism through geometrical symmetry breaking, thereby increasing the efficiency of amplitude modulation six-fold, and achieving modulation depths of 40% at ±30 V. This work demonstrates the potential of active and passive resonance control enabled by high-Q GMR structures for efficient electro-optic modulation or multifunctional sensing.

## Introduction

Optical light–matter interactions provide access to probing, controlling, and manipulating physical systems. Enhancing the sensitivity of such interactions is critically important for detecting weak perturbations and enabling efficient optical modulation at small scales. Resonant photonic systems offer a powerful platform for enhancing light–matter interactions through strong electromagnetic field confinement and narrowband resonances. Optical metasurfaces enable precise control of free-space electromagnetic waves through subwavelength engineering of optical resonances and wavefront responses. In particular, guided-mode resonance (GMR) metasurfaces[1-6], formed by periodically structured waveguides supporting high quality-factor (Q-factor) optical modes, exhibit strong spectral and phase responses to weak variations in geometry and refractive index, enabling highly sensitive beam manipulation and sensing.[7-19] Interactions between spectrally detuned resonances can further amplify optical responses, conceptually analogous to the Vernier effect employed in precision metrology and integrated photonic systems, where small relative resonance shifts produce strongly enhanced optical responses. Here, we enhance optical sensitivity through engineered high-Q guided-mode resonances exhibiting strong responses to geometric and refractive-index variations, enabling efficient electro-optic modulation in a transmissive metasurface. The resonant elements are formed by silicon waveguides with a well-defined perturbation period, patterned with 100 pm-scale precision on lithium niobate. The electro-optic substrate serves as the active medium, whose refractive index is modulated under externally applied electric fields via the Pockels effect.[20–28] By integrating interdigitated push–pull electrodes between neighboring GMR elements in the metasurface, we achieve periodic electro-optic detuning of the resonances across the array. This controlled symmetry breaking enables voltage-tunable diffraction and beam control via simultaneous phase and amplitude modulation of subwavelength-spaced radiative GMR elements. Precise geometric control enables the realization of spectrally detuned resonances that enhance the sensitivity of the optical response to electro-optic perturbations, analogous to Vernier-type amplification in coupled resonant systems. The architecture naturally supports cascaded optical systems, enabling integration with chip-scale light sources and detectors and allowing multiple functional layers to be stacked without obstructing the optical path.[29–31]

## Results

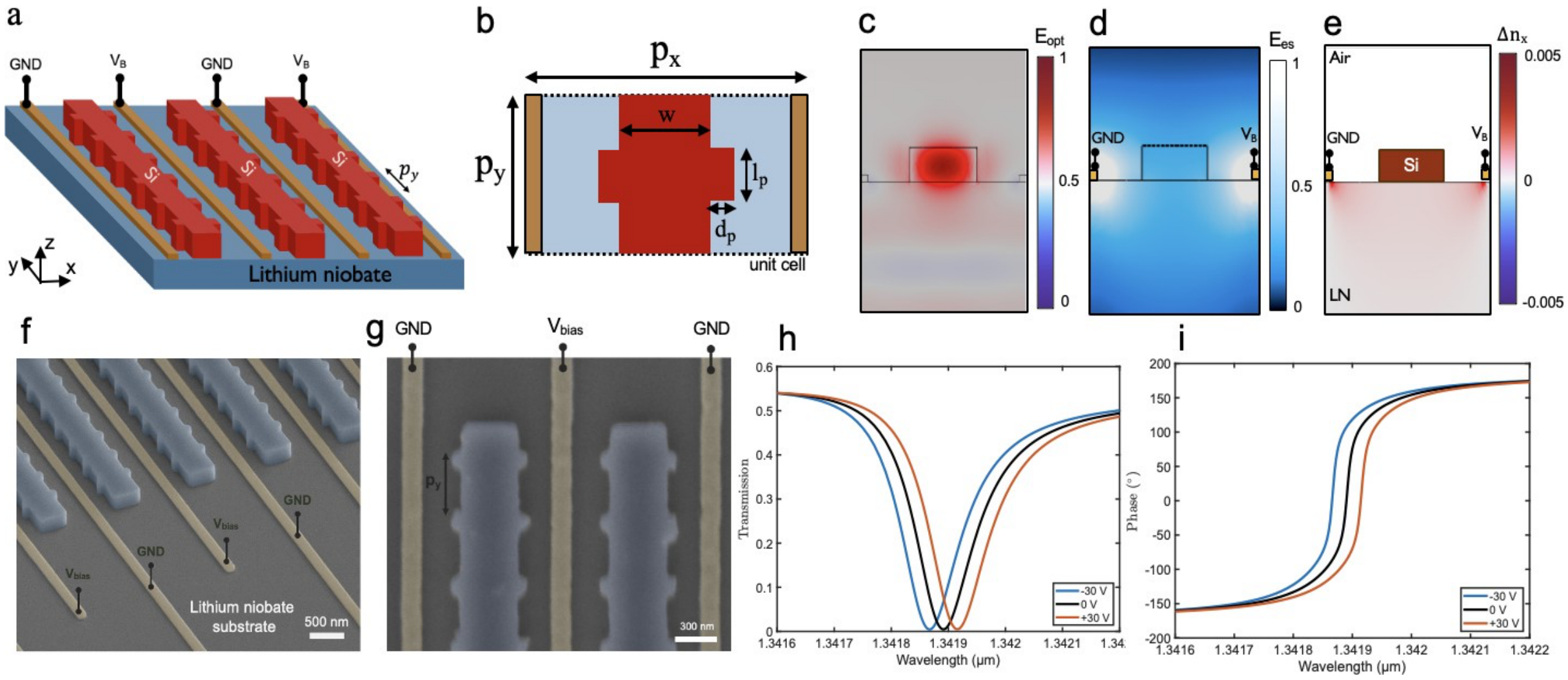

**Figure 1: Electro-optic lithium niobate transmissive metasurface.** (a) Schematic of the metasurface consisting of periodically perturbed silicon waveguides (supporting guided-mode resonances, GMRs) on a lithium niobate (LN) substrate. (b) Top view of a unit cell. (c) Simulated optical field $E_{opt}$ profile of the guided mode confined in a single perturbed waveguide, showing strong vertical confinement. (d) Simulated electrostatic field under applied voltage. (e) Resulting refractive index modulation $\Delta n_x$ in the lithium niobate induced by the applied voltage via the Pockels effect. (f) Tilted-view scanning electron microscope (SEM) image of the fabricated GMR metasurface. (g) SEM image of two adjacent GMR resonator elements with a single driving electrode positioned between them. (h) Guided-mode resonance in the transmission spectrum at 0 V and under applied bias of ±30 V. (i) Phase response extracted from simulations at 0 V and under applied bias ±30 V, revealing electro-optic tunability of the transmitted phase.

The proposed active transmissive metasurface consists of an array of silicon waveguides (w × h = 400 nm × 240 nm) fabricated on a bulk lithium niobate substrate (Figure 1). Each waveguide incorporates periodic sidewall perturbations with dimensions of $d_p$ = 50 nm and $l_p$ = 80 nm, enabling the excitation of guided-mode resonances with the Quality factor $Q = \lambda/\Delta\lambda$ at the free-space wavelength $\lambda$. The periodic perturbations provide the necessary momentum to couple light from free space into the waveguides, enabling the excitation of the resonant mode inside the waveguide (Figure 1c). The periodicity of the perturbation determines the spectral position of the guided mode resonance, according to the phase matching condition $k_0 \sin\theta + mG = k_0\, n_{\mathrm{eff}}$ with the free-space wave vector $k_0$, the angle of incidence $\theta$, the diffraction order $m$, the grating vector $G = 2\pi/\Lambda$ with period $\Lambda$ and the mode effective index $n_{\mathrm{eff}}$ of the guided mode. Driving electrodes are integrated to modulate the refractive index of the substrate by applying an electric field (Figure 1d–e). A change in the refractive index modifies the mode effective index $n_{\mathrm{eff}}$ which in turn spectrally shifts the guided-mode resonance. This shift is reflected in both the spectral position of the transmission dip and in the corresponding phase response (Fig. 1h–i) of light passing through the metasurface under normal incidence ($\theta = 0°$). A phase shift as high as 200 degrees can be achieved by applying a voltage of 30 V.

The device was fabricated using conventional lithographic nanofabrication processes. A 240-nm-thick amorphous silicon film was structured on a bulk x-cut lithium niobate substrate (see the "Methods" section for more details). Arrays of 150 $\mu$m × 150 $\mu$m silicon GMR elements with a well-defined perturbation period were fabricated to allow deterministic tuning of the spectral position of the guided-mode resonance. Interdigitated electrodes were integrated adjacent to the waveguides to enable push–pull electro-optic tuning.[26,27,32]

The metasurface was oriented such that the electrodes are aligned perpendicular to the optical axis of the lithium niobate crystal, thereby maximizing the overlap with its largest Pockels coefficient. $r_{33}$ = 32 pm/V. Precise alignment between the lithographic steps is required to ensure that the electrodes (width × height = 90 nm × 50 nm) are positioned centrally between the silicon GMR elements, minimizing the impact of optical loss from the metallic electrodes on the guided-mode resonance (Supplementary Note 1).

To experimentally characterize the metasurface, we illuminated the device with linearly polarized light at normal incidence through the substrate, using a supercontinuum light source or a wavelength-tunable diode laser. The transmitted light was collected with an objective lens and routed either to an InGaAs infrared camera for imaging or focused onto a photodetector for electro-optic characterization. A pronounced dip in the transmission spectrum appears at resonance, exhibiting a minimum full width at half maximum (FWHM) linewidth of 0.5 nm (Figure 2). The minimum transmission at resonance ranges between 20–40%. Fitting the spectral response with a Lorentzian lineshape function reveals GMR Q-factors with values as high as 3000 (Supplementary Note 2). As the perturbation period decreases, the resonance dip shifts toward shorter wavelengths, in agreement with the expected phase-matching condition of a guided-mode resonance. We further investigate electro-optic modulation in a metasurface structure with a uniform perturbation period of $p_y$ = 490 nm, which exhibits a guided-mode resonance at 1283 nm with a measured Q-factor of 2400. Since the GMR condition depends critically on the effective refractive index of the silicon waveguides, even small fabrication-induced variations in the silicon refractive index or waveguide geometry can result in pronounced shifts of the resonance wavelength (Supplementary Note 3). This sensitivity readily explains the observed deviation of the GMR spectral position from simulations based on the nominal geometry and literature values of the optical constants. To account for fabrication uncertainties, we refined the waveguide geometry and the optical constants of silicon in our model within realistic tolerance ranges, based on SEM and AFM characterization of the fabricated devices, to achieve agreement with the experimentally observed spectral position of the GMR resonance (Supplementary Note 4). While the measured quality factor is high ($Q_{exp}$ = 2400), it is lower than the simulated value ($Q_{sim}$ = 8350) due to the idealized lossless material properties and perfectly periodic and infinite geometry assumed in the simulations. To account for material absorption, we introduce a finite optical loss $k_{Si}$ = 0.0001, which contributes to the intrinsic damping of the resonant mode and leads to a reduced simulated quality factor in closer agreement with experiment. The adjusted parameters are retained for all subsequent simulations and analysis presented in this work.

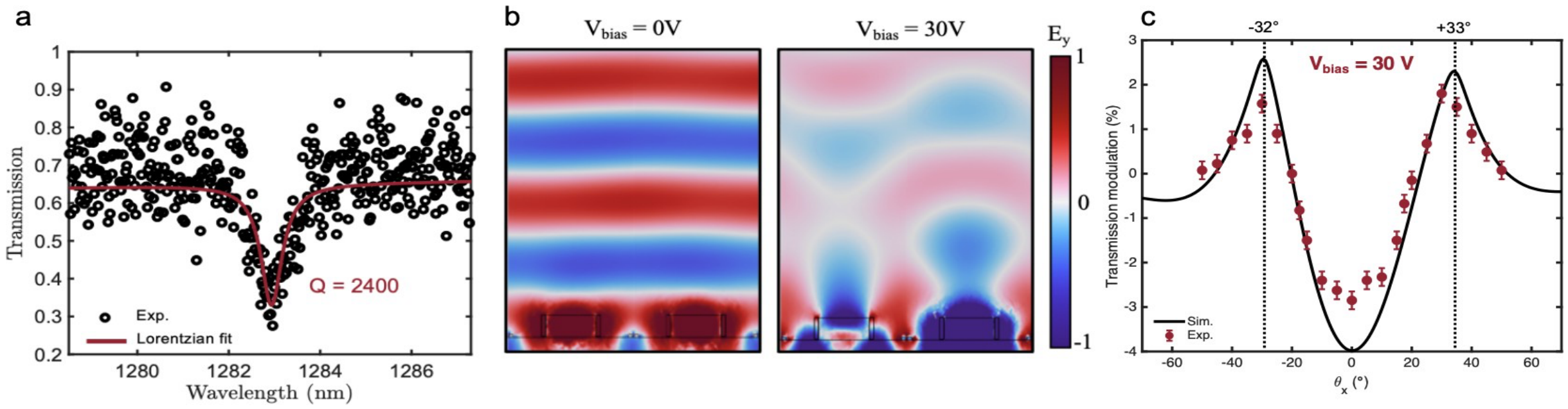

**Figure 2: Symmetric guided-mode-resonance metasurface enabling electro-optic beam splitting via dynamic diffraction**. (a) Transmission spectrum showing the guided-mode resonance at $\lambda$ = 1283 nm with a quality factor of $Q$ = 2400 (b) Simulated near-field distribution of the out-of-plane electric field component $E_y$ under zero bias (left) and an applied bias of 30 V (right), illustrating the emergence of spatially periodic radiation under electro-optic actuation. (c) Measured modulation depth as a function of in-plane wavevector $k_x$, demonstrating voltage-controlled redistribution of optical power into the first diffraction order.

When a bias is applied across the interdigitated push–pull electrodes, the electric field direction alternates beneath adjacent GMR elements. Owing to the linear electro-optic effect in lithium niobate, this produces refractive-index changes of opposite sign via the $r_{33}$ tensor component, resulting in opposite resonance shifts between neighboring resonator elements. These shifts introduce phase modulation while leaving the radiative coupling strength and spectral resonance linewidth unchanged. From the expansion of the transmission, $T(\lambda \pm \Delta\lambda) \approx T(\lambda) \pm \Delta\lambda\delta\lambda T(\lambda)$, the opposite resonance shifts lead to a cancellation of the linear terms, such that the total transmitted amplitude remains constant to first order. Consequently, the resulting electro-optic modulation introduces phase modulation with only small global amplitude variation. The metasurface is designed with subwavelength spacing between adjacent GMR elements and therefore supports only the zeroth diffraction order under zero bias, in the absence of any electro-optically induced symmetry breaking between elements. Under push–pull electro-optic tuning, antisymmetric phase modulation across a two-element unit cell forms a dynamic grating with an effective period larger than the operating wavelength, thereby introducing electro-optically controlled diffraction orders under applied bias.

The subwavelength radiative elements experience both phase and amplitude modulation, which together determine the diffraction efficiency of the electro-optic grating. Maximum diffraction efficiency is achieved for local phase-only radiative elements. However, as shown in Fig. 1h and Fig. 2a, the transmission within each radiative element varies strongly around the guided-mode resonance, with the transmission minimum at the resonance wavelength. Consequently, the electro-optic modulation intrinsically couples local phase and amplitude variations, leading to deviations from the ideal phase-only diffraction. Although the phase response is strongest at the guided-mode resonance, the simulated local transmission minimizes at this spectral position. Optimal device operation therefore occurs on resonance, where a phase modulation is maintained while sufficient transmission is enabled by a symmetric shift of the transmission. We estimate the expected diffraction efficiencies from the voltage-dependent amplitude and phase profiles obtained by FEM simulations (Figure 2b). A near-field to far-field transformation is applied to compute the diffracted power distribution. The transmitted power at the respective wavelength is used for normalization of the electro-optic (EO) diffraction efficiencies. The diffraction angles are determined by the electro-optically induced grating period, $d_{EO}$ = 2d. With the nominal separation d = 1200 nm of GMR elements, the push–pull modulation creates an effective supercell of $d_{EO}$ = 2400nm that sets the diffraction angle to $\theta_x = \pm(32 \pm 1)°$ (Figure 2c). Using Fourier-plane imaging of the transmitted field, we quantify the modulation depth in momentum space and observe electro-optic diffraction at an operating wavelength of 1283.1 nm. The maximum diffraction efficiency is (3.2±0.5) %.The experimentally measured diffraction efficiencies agree with the simulated values, with minor deviations attributable to fabrication tolerances and material losses.

To enable efficient electro-optic amplitude modulation in transmission, we implement a second metasurface design in which neighboring GMR elements are passively detuned (Figure 3), thereby enhancing the sensitivity of the optical response to electro-optic refractive index changes and enabling larger modulation depths through enhanced resonance interactions.This is achieved by introducing a slight offset in the perturbation period $p_y$ between every second GMR element. As a result, the unit cell comprises two structurally distinct radiative elements that exhibit individually different resonance wavelengths. Notably, this detuning is introduced structurally through sub-nanometer variations in the perturbation period. The GMR response is governed by the collective lattice resonance of an extended array of unit cells (~300 elements spanning the 150-μm-long metasurface), such that the effective detuning is defined by the ensemble response of the entire resonator array. Consequently, the effective resonance condition reflects an ensemble-averaged structural perturbation that is statistically well defined over hundreds of elements, enabling reproducible and robust sub-nm pertubation shifts despite the finite accuracy of the used nanofabrication processes. We experimentally resolve resonance shifts as small as 0.45 nm, arising from a $\Delta p_y$ = 400 pm variation in the perturbation period of the metasurface. This shift is slightly smaller than predicted by simulations (Fig. 3c), which we attribute to the extreme sensitivity of the resonance to nanometer-scale geometric variations and uncertainties in the optical properties of the metasurface materials.

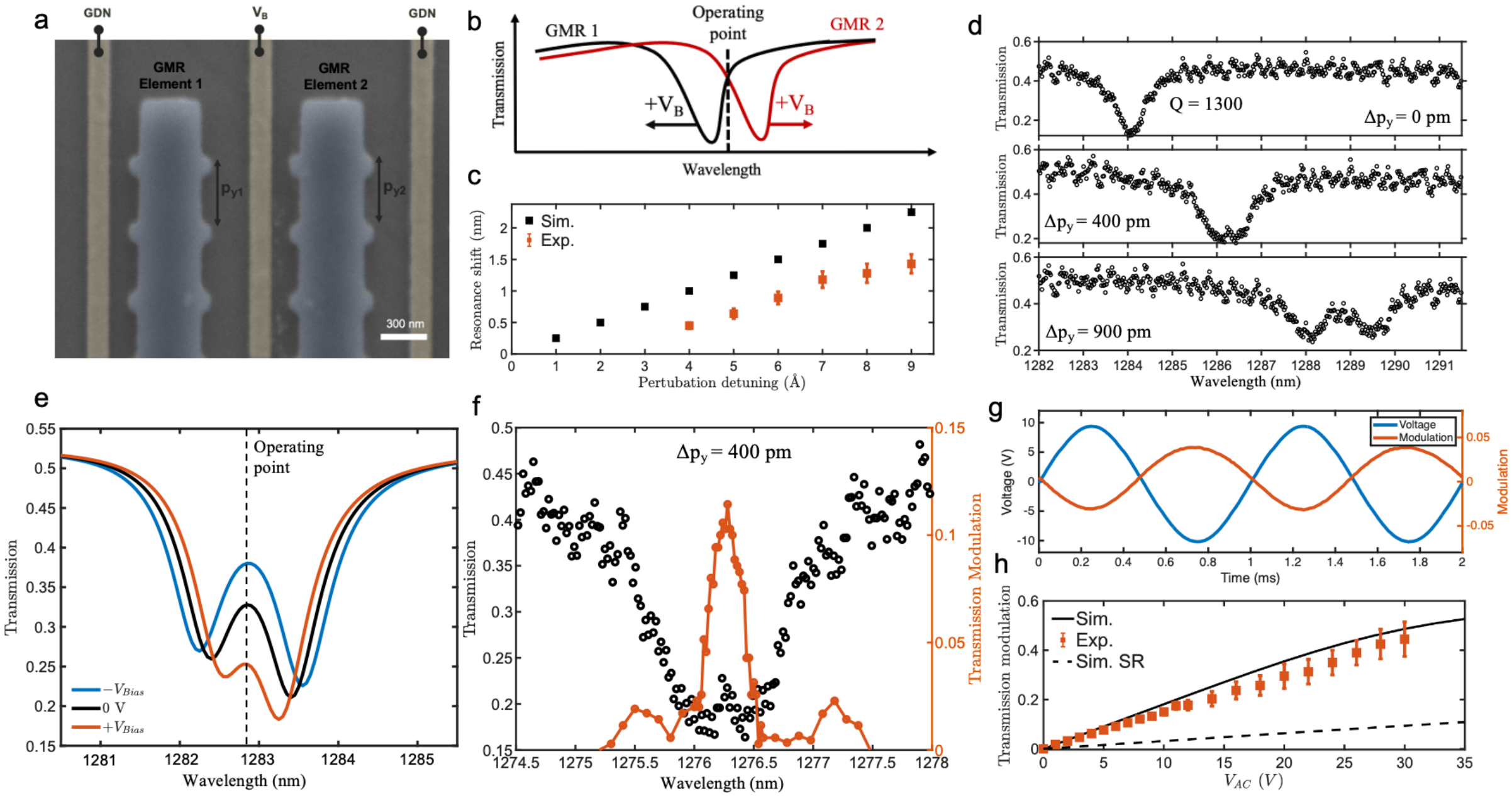

**Figure 3: Detuned guided-mode-resonance (GMR) Venier metasurface**. (a) SEM image of two neighboring guided-mode-resonance (GMR) elements with slightly different perturbation periods $p_{y1}$ and $p_{y2}$, forming a detuned two-element unit cell with interdigitated push–pull electrodes. (b) Schematic illustration of two spectrally detuned GMRs and the operating point between them, where opposite electro-optic resonance shifts under applied bias lead to asymmetric transmission modulation. (c) Simulated and experimental resonance shift as a function of the perturbation-period offset. (d) Spectral response of metasurfaces with geometrical detuning parameters $\Delta p_y = 0$ pm, $\Delta p_y = 400$ pm, and $\Delta p_y = 900$ pm. (e) Simulated transmission spectrum of the detuned GMR metasurface under an applied bias of $V_{bias} = \pm 15$ V, demonstrating pronounced amplitude modulation between the guided-mode resonances. An optical loss of $k_{Si} = 0.0005$ was introduced for silicon to obtain a resonance quality factor of $Q = 1300$, matching the experimentally observed value. (f) Transmission modulation spectrum for $\Delta p_y = 0.4$ at a driving voltage of $V_{AC} = 10$ V. (g) Time-domain measurement of electrical drive voltage and resulting optical transmission modulation, confirming linear electro-optic response at the operation point. (h) Modulation depth as a function of the driving voltage $V_{AC}$, comparing simulation results for the detuned metasurface (solid line) and the single-resonance (SR) metasurface (dashed line) with experimental data for the detuned metasurface.

When an electrostatic field is applied using the push–pull electrodes, the resonance wavelengths of the two detuned elements shift in opposite directions, producing a strong modulation of the transmitted amplitude at an intermediate wavelength between the two overlapping GMR resonances. Unlike the symmetric device, in which antisymmetric resonance shifts predominantly induce phase modulation while preserving the overall transmission, the intentional detuning breaks this cancellation when the device is operated at a wavelength positioned between the slightly overlapping resonances. Consequently, the opposite spectral shifts give rise to an asymmetric transmission response within the two-element unit cell, resulting in strongly enhanced amplitude modulation of the transmitted field through a response analogous to the Vernier effect. A maximum transmission modulation of 40% is achieved at a driving voltage of 30 $V_{AC}$ in the 400 pm-detuned metasurface. For the experimentally-observed GMR with Q = 1300, the optimum detuning of $\Delta p_y$ = 400 pm results in a six-fold enhancement of the amplitude modulation efficiency (Supplementary Note 5). Notably, the modulation depth can be further increased by exploiting higher-Q resonances through reduced scattering and optical losses in the metasurface. Although the absolute operating wavelength may drift due to temperature fluctuations or mechanical deformation, which were neither observed nor explored in the present experiments, both resonances are expected to shift synchronously, thereby preserving their relative detuning. As a result, global wavelength drifts shift both resonances together without altering their separation, preserving the operating point between them and ensuring inherent robustness against spectral fluctuations. This characteristic makes the passively detuned metasurface particularly attractive for ultrasensitive sensing applications,[33–36] as well as for stable electro-optic amplitude modulation in environments subject to thermal or refractive-index fluctuations, or in systems where the operating wavelength is intentionally tuned, such as thermally controlled distributed-feedback (DFB) lasers. Due to the inherent nature of active transmissive metasurfaces, cascaded operation and inherent integration with chip-scale lasers are straightforward, enabling compact and scalable photonic system architectures.[37,38]

## Conclusion

We demonstrate an active transmissive metasurface based on high-Q guided-mode resonances in silicon-on-lithium niobate that enables electrically controlled phase and amplitude modulation through symmetry tuning. Interdigitated push–pull electrodes induce opposite electro-optic resonance shifts in neighboring GMR elements, resulting in electro-optic diffraction in geometrically symmetric metasurfaces. Introducing 100 pm scale passive detuning between adjacent elements intentionally breaks this cancellation, converting antisymmetric resonance shifts enabling a sixfold increase in transmission amplitude modulation efficiency. As the response arises from the collective behavior of a large ensemble of unit cells, the modulation exhibits intrinsic tolerance to fabrication variations and thermal drift. These results establish a general strategy for combining passive resonance control with active electro-optic tuning in transmissive metasurfaces, enabling compact, cascaded photonic platforms for dynamic beam shaping, modulation, and sensing.

## Methods

### Numerical modeling

Finite element method (FEM) simulations are performed using a commercially available software (Comsol Multiphysics 6.2). For the electrostatic simulations, the device unit cell is modeled with the unclamped static relative permittivity tensor of LN taken from Jazbinsek *et al.*[34] ($\epsilon_{xx} = \epsilon_{yy} = 27.8$, $\epsilon_{zz} = 84.5$), while the relative permittivity of air is set to be $\epsilon_{air} = 1$, and the boundaries of the two gold nanostripes are set to ground and $V_{bias}$ potential, respectively. The calculated electric-field distribution is utilized to determine the modification of the refractive index in the LN substrate by using the electro-optic Pockels coefficients from Jazbinsek *et.al.*[34] (restricting to the largest diagonal terms, i.e., $\Delta n_{ii} = -0.5 r_{iiz} n_{ii3} E_x$, with $r_{yyx} = r_{zzx} = 10.12$pm/V and $r_{xxx} = 31.45$pm/V), which are considered to be non-dispersive over the investigated wavelength range. For the electro-optical simulation, the modified distribution of the refractive index of LN is fed into the 3D full-wave simulation, while the unmodified refractive indices from Au and LN were taken from Johnson and Christy[39] and Zelmon *et al.*[40] ($n_{xx} = n_{yy} = n_o = 2.221$, $n_{zz} = n_e = 2.146$ at $\lambda_0 = 1280$nm). Excitation and detection ports in combination with periodic boundary conditions at the sidewalls of the unit cell are applied to calculate the transmission and phase as a function of the applied voltage, which is used for investigating the shift of the GMR and the corresponding transmission and phase response.

### Device fabrication

Devices are fabricated on commercially available x-cut lithium niobate substrates with gold alignment markers, which are defined in a prior step using electron beam lithography, metal deposition, and lift-off. The silicon waveguide array metasurfaces are patterned by electron beam lithography (Raith EBPG 5200) in a spin-coated 200-nm-thick PMMA A4 positive resist. The electron-beam exposure uses a minimum beam step size of 0.1 nm. After development, a 15 nm layer of $Al_2O_3$ is deposited to serve as an etch hard mask, followed by lift-off. A $SF_6/C_4F_8$ etch chemistry is used to transfer the pattern into the silicon over a 2-minute etch. The gold alignment markers are used for lithographic overlay in the subsequent electron beam lithography step, where the electrodes (metasurface driving electrodes and macroscopic bonding pads) are defined. The chip is wire-bonded onto a printed circuit board (PCB), enabling electrical interfacing with external driving.

### Electro-optic characterization

The metasurfaces were characterized using a home-built optical transmission microscope.

Coherent light from a wavelength-tunable diode laser (Santec TSL-510) or a broadband light source was loosely focused on the metasurface. The transmitted light was collected with an objective lens (Zeiss EC Epiplan-Apochromat 20x/0.6) and projected either onto an InGaAs IR camera (Xenics Bobcat 320) or a power meter (Thorlabs DET10C). For power normalization, the sample was removed, and the illuminated power was measured through the same area. For the beam deflection measurements, a 0.9 NA objective was used to image the Fourier plane onto the camera.

**Acknowledgement**

This work was performed in part at the Kavli Nanoscience Institute at the California Institute of Technology. H.A.A. acknowledges support from the Air Force Office of Scientific Research Meta-Imaging MURI Grant No. FA9550-21-1-0312 and Sony Corporation.

**Supporting Information**

Supplementary Note 1: Effect of Electrode Offset on Resonance Linewidth and Quality Factor, Supplementary Note 2: Spectral GMR Shift Induced by Variation of the Perturbation Period, Supplementary Note 3: Spectral GMR Shift Induced by Variation of the Optical Constamts and Metasurface Geometry, Supplementary Note 4: Calibration of Silicon Refractive Index and Waveguide Geometry. Supplementary Note 5: Modulation enhancement in detuned GMR resonators